\documentclass[runningheads]{llncs}
\usepackage[T1]{fontenc}
\usepackage{graphicx}
\usepackage{booktabs}
\usepackage{caption}
\usepackage{amssymb}
\usepackage{subcaption}
\usepackage[misc]{ifsym}
\usepackage{amsmath}
\usepackage{multirow}
\newcommand{\corr}{(\Letter)}
\usepackage{algorithm}
\usepackage{algorithmic}
\usepackage{svg}  
\usepackage{ragged2e}

\usepackage{mwe}

\begin{document}

\title{Heterogeneous Sequel-Aware Graph Neural Networks for Sequential Learning}

\titlerunning{HSAL-GNN}

\author{Anushka Tiwari\inst{1} \and
Haimonti Dutta\inst{1,2}\corr \and
Shahrzad Khanizadeh\inst{2}}


\institute{Institute for Artificial Intelligence and Data Science, \\
The State University of New York at Buffalo \\
\email{\{atiwari4\}@buffalo.edu}
\and
Department of Management Science and Systems, \\
The State University of New York at Buffalo \\
\email{\{haimonti,skhaniza\}@buffalo.edu}
}

\maketitle              

\begin{abstract}
Graph-based recommendation systems use higher-order user and item embeddings for next-item predictions. Dynamically adding collaborative signals from neighbors helps to make use of preferences of similar users during learning. While item-item correlations and their impact on recommendations have been studied, the efficacy of \emph{temporal item sequences} for recommendation is much less explored. In this paper, we study temporal item sequence (sequel-aware) embeddings along with higher-order user embeddings and show that sequel-aware Graph Neural Networks have better (or comparable) recommendation performance than graph-based recommendation systems that do not take sequel information into consideration. 
Extensive empirical results comparing Heterogeneous Sequel-aware Graph Neural Networks (HSAL-GNNs) to other algorithms for sequential learning (such as transformers, graph neural networks, auto-encoders) are presented on three synthetic and three real-world datasets. Our results indicate that incorporation of sequence information from items greatly enhances recommendation. 

\end{abstract}

\section{Introduction}
Sequential recommendation refers to next-item prediction, given historical records of a user's interaction with a set of items, and the evolution of their preferences over time \cite{Zhang_23a,Li_23a}. Early work in this problem domain involved modeling user-item interactions on static graphs, but more recently emphasis has shifted to capturing dynamically evolving user-preferences. Methods such as Markov Chains \cite{Rendle_10a}, RNN-based methods (such as LSTMs \cite{Sak_14a} and Gated Recurrent Unit (GRU) Networks \cite{Chung_14a}), Convolutional Neural Networks \cite{Tang_18a}, Attention Models \cite{Kang_18a} and Graph Neural Networks \cite{Zhang_23a} have been applied to model evolving user-preferences which in turn enables next-item prediction.

Attempts to model \emph{item} representations have also received some attention. Variational Auto Encoders (VAEs) have been used to obtain representation distributions of items preferred by users and inject uncertainty into them (\cite{Liang_18a},\cite{Sachdeva_19a},\cite{Xie_21a}). Li et al. \cite{Li_23a} proposed to model items as distributions which reflect a user's multiple interests and item features. Fan et al. \cite{Fan_21a} proposed modeling the item representation as elliptical Gaussian distributions (DT4SR) which serves to reveal a user's basic interests, and a stochastic vector  introduced for covariance representation and uncertainty injection. 
This method was further modified \cite{Fan_22} to include a novel stochastic self-attention (STOSA) model which replaces the inner product between any of two items in a self-attention module as the Wasserstein distance for measuring correlation and transitivity. 

Items (such as books, movies) often have sequels. For example, \textit{Avatar}, an American epic science fiction created by James Cameron has five sequels -- the first was released in December, 2009, the second called \textit{Avatar: The Way of Water} was released in December, 2022; the third \textit{Avatar: Fire and Ash} will be released in 2025 and Avatar 4 and 5 have been planned for releases in 2029 and 2031 respectively. Quoting an article about this film \cite{Avatar}, 
``Like the original film, the four sequels have \textit{fully encapsulated} stand-alone plots that ``come to their own conclusions". The four films have an overarching meta-narrative that connects them to create a large interconnected saga. Cameron described the sequels as ``a natural extension of all the themes, and the characters, and the spiritual undercurrents" of the first film."
 While such natural extensions for books, movies, and other items are common in the real-world (for e.g. James Bond has $23+$ movies since 1962, Godzilla has $28+$ movies since 1954; books such as Harry Potter has seven sequels leading to eight movies in chronological order) rarely have the temporal sequences been used for item representation. To the best of our knowledge, our work is the first work that captures temporal information from item sequences, in addition to dynamic user-preference modeling. 
 
 Our research focuses on the fact that a user may purchase some standalone (non-sequential) items along with items with sequels. The primary research questions we address in this paper are: (1)  How does the presence of sequels affect sequential item recommendation? (2) If a user has $k$ historical interactions, some of which pertain to sequential items and others non-sequential, how does the mix affect overall sequential item recommendation?

This paper is organized as follows: Section~\ref{related} presents related work; Section~\ref{prelim} presents notation and preliminaries; Section~\ref{HSAL-GNN} presents our main algorithm; Section~\ref{exp} presents empirical analysis, and Section~\ref{conc} concludes the work.

\section{Related Work}
\label{related}

\subsection{Sequential Recommendation}
In the early years of research on sequential recommendation systems, Markov Decision Processes (\cite{Rendle_10a}, \cite{Shani_05a}) were used extensively to model user-item interactions. 
With the popularity of deep sequential networks, CNN, GRU, LTSM and their variants are found extensively in the literature. GRU4Rec \cite{Hidasi_2015} stacks layers of GRU and makes use of mini-batch training for sequential recommendation. SASRec \cite{Kang_18a} follows an autoregressive model and uses unidirectional self-attention mask to capture previous item information for prediction of the next item. BERT4Rec \cite{Sun_19a} uses a bidirectional self-attention mechanism arguing that future items may also influence current item prediction. 

It is well known however, that GNNs are best known for their ability to capture higher order user-item information. 
Ding et al. \cite{Ding_21a} model user-item interaction and item-item transitions to obtain enhanced user and item representations by incorporating higher-order connections over the graphs.
Wu et al. \cite{Wu_19a} study session-based recommendations wherein sequences are modeled as graph-structured data. From the session graph, GNNs are used to capture complex transitions of items. Each session is then represented as the composition of the global preference and the current interest of that session using an attention network.

\subsection{Modeling Item Representations}
To learn item distributions, transformer-based approaches such as Distribution Transformer for Sequential Recommendation (DT4SR)\cite{Fan_21a} and STOchastic Self-Attention (STOSA) \cite{Fan_22} are known in the literature.  
DT4SR uses elliptical Gaussian distributions to describe item sequences with uncertainty, which also helps alleviate cold start problems. Wasserstein distance measures the similarity between distributions. STOSA, on the other hand, embeds each item as a stochastic Gaussian distribution whose covariance encodes uncertainty. A regularization term used alongside the ranking loss ensures that dissimilarity between positive and negative items is maintained. Variational Auto Encoders (VAE)s are able to model probabilistic latent variables and often used to inject uncertainty in sequential recommendation. Liang et al. \cite{Liang_18a} introduce a generative model with multinomial likelihood and Bayesian inference for parameter estimation. It has been shown that the resulting learning algorithm has connections to maximum entropy discrimination. Sachdeva et al. \cite{Sachdeva_19a} introduce a recurrent VAE, where they pass a subset of the history of user-item preferences through a Recurrent Neural Network (RNN). At each time-step of the RNN, the sequence is fed through a series of fully-connected layers, the output of which models the probability distribution of the most likely future preferences. It has been shown that handling temporal information in this manner is crucial for improving the accuracy of the VAE. However, VAE-based methods suffer from representation degeneration and model collapse issues. 

\section{Preliminaries}
\label{prelim}


\subsection{Sequel-Aware Sequential Recommendation} In this section, we introduce the problem of sequential recommendation with structured sequel dependencies and define key notations.
In sequential recommendation, let \( \mathcal{U} \) and \( \mathcal{I} \) denote the set of users and items, respectively. Each user \( u \in \mathcal{U} \) interacts with a sequence of items over time, forming an interaction sequence \( S^u = (i_1, i_2, \dots, i_k) \) with the corresponding timestamps \( T^u = (t_1, t_2, \dots, t_k) \). The items may be categorized into standalone items $\mathcal{I}_{SA}$, which do not belong to any sequence\footnote{The corresponding user-item interaction sequence is represented as $S_u^{SA}$}, and items with sequel $\mathcal{I}_{SQ}$, which are part of an ordered sequence\footnote{The corresponding user-item sequence is represented as $S_u^{SQ}$}. Examples of items in sequence can be books or movies (e.g., $ \text{Harry Potter}1 \rightarrow \text{Harry Potter}2 \rightarrow \text{Harry Potter}3$). The objective of this research is to predict the next item by leveraging both user-item (standalone and sequential) interaction history and \emph{structured relationships for items with sequence}. 

Each user and item is represented by a low-dimensional embedding vector \( \mathbf{e}_u \in \mathbf{R^d} \), \( \mathbf{e}_i \in \mathbf{R^d} \), where \( d \) is the embedding dimension. The user embedding matrix is denoted as \( \mathbf{E_U} \in \mathbf{R}^{|\mathcal{U}| \times d} \) and the item embedding matrix as \( \mathbf{E_I} \in \mathbf{R}^{|\mathcal{I}| \times d} \). To encode the relative position of an item within its sequence, we introduce a positional embedding function \( P(i) \) computed as:

\begin{equation}
P(i)_{(2k)} = \sin \left( \frac{p_i}{10000^{2k/d}} \right), \quad P(i)_{(2k+1)} = \cos \left( \frac{p_i}{10000^{2k/d}} \right),
\end{equation}
where \( p_i \) represents the position of item \( i \) within its sequence, \( d \) is the embedding dimension, and \( k \) indexes the embedding vector dimensions.

\subsection{Sequel-Aware Dynamic Graph}
To model both user-item interactions and structured sequel dependencies, we construct a sequel-aware graph \( G \) that distinguishes between standalone and sequel items. 
A sequel-aware dynamic graph is formally defined as:
\begin{equation}
G = (\mathcal{U}, \mathcal{I}_{SA} \cup \mathcal{I}_{SQ}, \mathcal{E}_U, \mathcal{E}_S, \mathcal{T}),
\end{equation}
where \( \mathcal{E}_U \) represents user-item interactions, capturing user engagement with items over time $\mathcal{T}$, while \( \mathcal{E}_S \) encodes sequel-aware item-item relationships to preserve structured dependencies. The graph $G$ is heterogeneous, as it contains two types of nodes -- users or items and edges encapsulate different kinds of interactions (user-item or item-item). 

Each node (user or item) in the graph is associated with an embedding, obtained from a Graph Neural Network (GNN) \cite{Scarselli_09a} built on it, and refined over multiple layers of message passing which we discuss later in the section~\ref{HSAL-GNN}. The representation of an item \( i \) at layer \( l \) is denoted by:
    $h_i^{(l)} \in \mathbf{R}^d, i \in \{\mathcal{I}_{SA},\mathcal{I}_{SQ}\}$,
and $d$  is the embedding dimension.


\section{Heterogeneous Sequel-Aware Graph Neural Networks for Sequential Learning (HSAL-GNN)}
\label{HSAL-GNN}
The heterogeneous sequel-aware GNN for sequential learning comprises of the following subparts: (a) Sequel-aware graph construction (b) Sub-graph sampling (c) GNN construction using message-passing, node aggregation, recommendation and optimization. These subparts are discussed at length in the following sub-sections. Schematic descriptions of the algorithms are provided in Appendix A.

\begin{figure}
    \centering
    \begin{subfigure}[b]{0.48\textwidth}
         \centering
         \includegraphics[width=\textwidth]{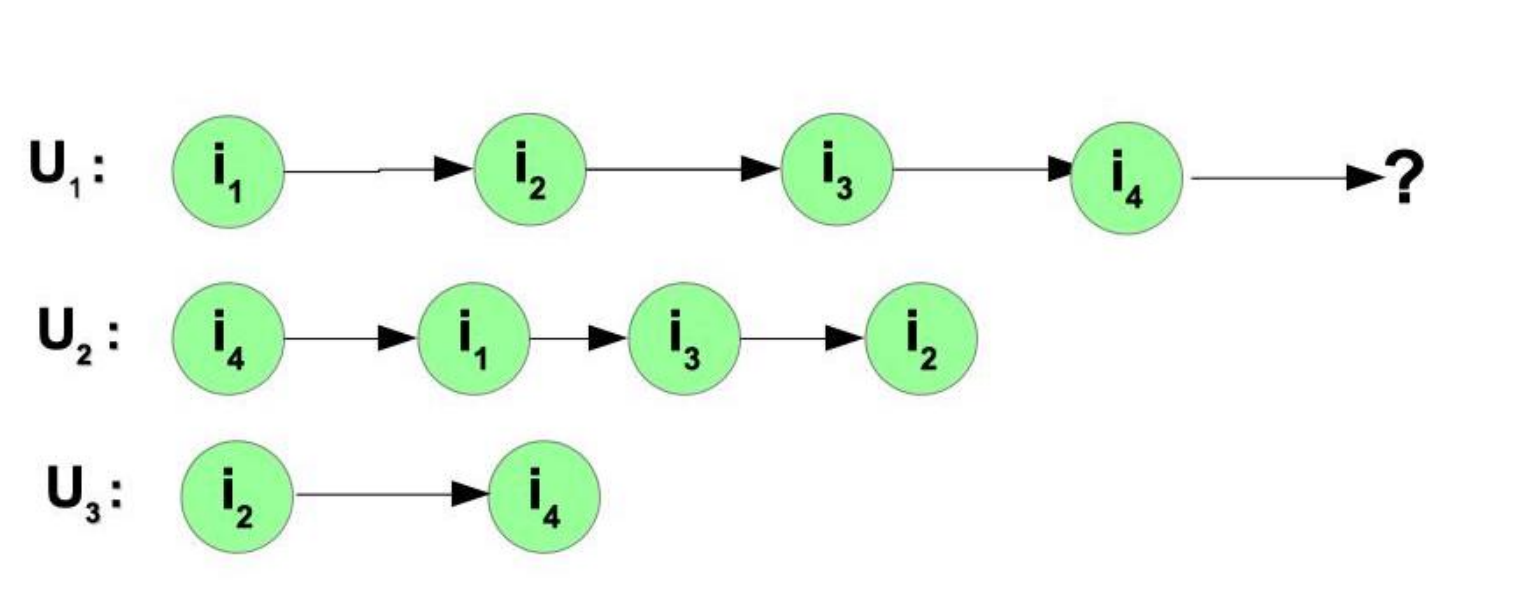}
     \end{subfigure}
     \hfill
     \begin{subfigure}[b]{0.48\textwidth}
         \centering
         \includegraphics[width=\textwidth]{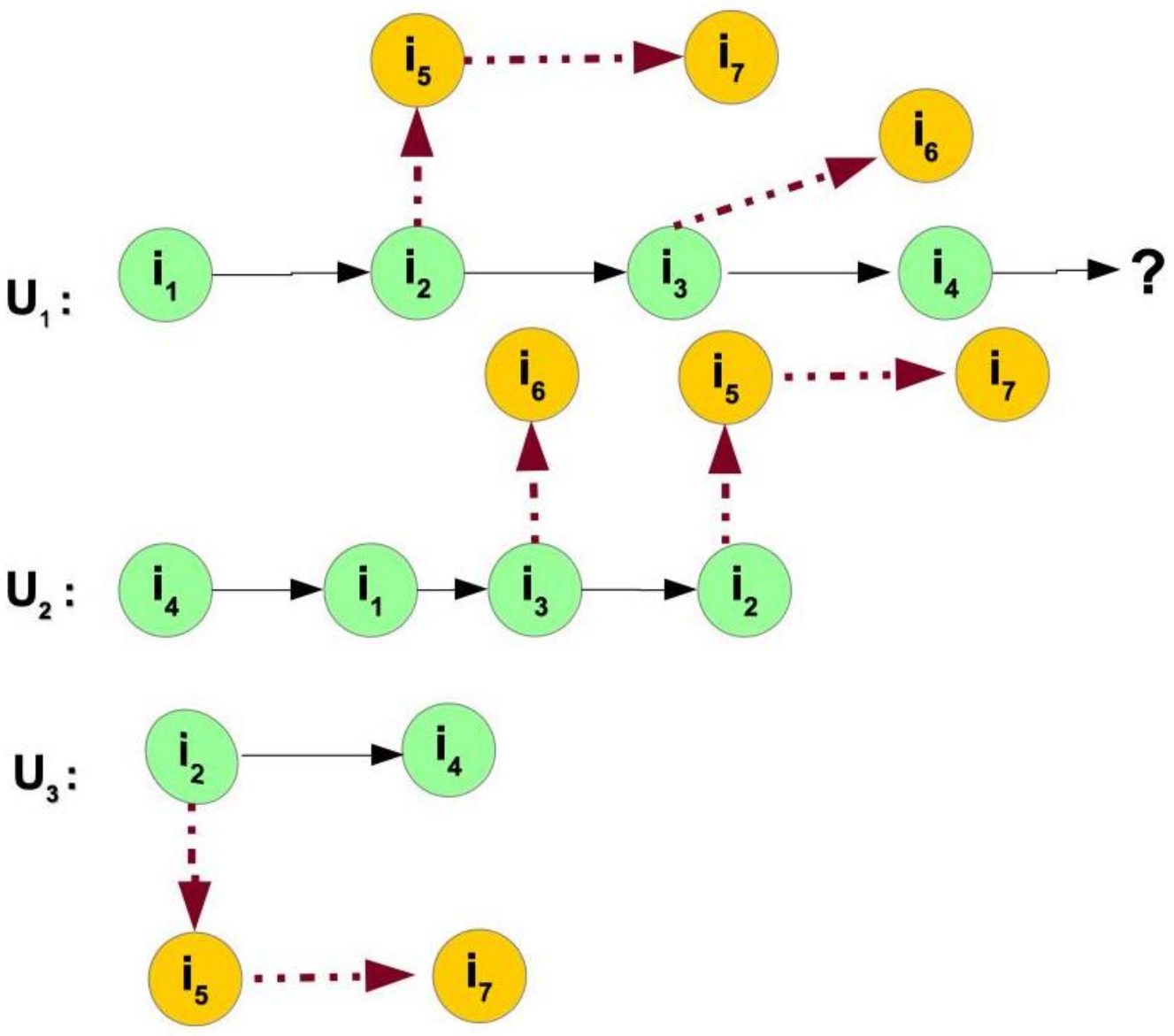}
     \end{subfigure}
     \hfill
    \caption{Illustrative Example $1$. (Left Figure) A sequential recommendation algorithm from literature.The goal is to predict the next item after $i_4$, using user-item interaction history. (Right Figure) A sequel-aware sequential recommendation problem. The goal is to predict the next item after $i_4$ using user-item interaction history and sequel information. The green nodes represent items a user has interacted with while the yellow nodes represent sequels. Several different edges are  seen - user-$i_{SA}$, user-${i_{SQ}}$, $i_{SA}$-$i_{SA}$, $i_{SQ}$-$i_{SQ}$, $i_{SA}$-$i_{SQ}$.}
    \label{illex1}
\end{figure}

\subsection{Sequel-Aware Graph Construction}
Given a user sequence \( S^u = (i_1, i_2, i_3, i_4) \) with timestamps \( T^u = (t_1, t_2, t_3, t_4) \) where \( t_1 < t_2 < t_3 < t_4 \), we first determine which items have sequels. Let $i_p, 1 \le p \le k$ be an item in $S^u$ such that $i_a$ and $i_b$ are items preceding and succeeding item $i_p$.
To capture relationships among sequels, we establish \emph{sequel-aware} edges between items as follows:
\begin{equation}
(i_a, i_p, s, p_a), \quad (i_p, i_b, s, p_b),
\end{equation}
where $s$ is the sequence identifier, and $p_a, p_b$ represent the relative positions of \( i_a \) and \( i_b \) in the sequel series. We also form user-item edges to distinguish interactions based on whether an item belongs to \( I_{SQ} \) or \( I_{SA} \). A user-item interaction is represented as:
\begin{equation}
(u, i_{sa}, t, p_{iu}, p_{ui}) \quad \text{if } i_{sa} \in I_{SA},
\end{equation}
\begin{equation}
(u, i_{sq}, t, p_{iu}, p_{ui}) \quad \text{if } i_{sq} \in I_{SQ}.
\end{equation}
Here, \( p_{iu} \) denotes the position of item \( i \) in the sequence of items interacted with by user \( u \), and \( p_{ui} \) represents the position of \( u \) among all users who interacted with \( i \). \\

\noindent{\textbf{An illustrative example (Figure~\ref{illex1}): }} Consider a user \( u_1 \) with $S^{u_1} =(i_1, i_2, i_3, i_4)$
at timestamps $T^{u_1} = (t_1, t_2, t_3, t_4)$, where \( t_1 < t_2 < t_3 < t_4 \).  
The standalone items are \( S_{u_1}^{\text{SA}} = (i_1, i_4) \), and the items with sequels are  
\( S_{u_1}^{\text{SQ}} = (i_2, i_3) \). Furthermore, let \( i_2 \) belong to sequence $A = (i_2, i_5, i_7)$, and \( i_3 \) belong to sequence $B= (i_3, i_6) $.  
The sequel-based dependencies form edges:
\[
(i_2, i_5, A, 2), \quad (i_5, i_7, A, 3), \quad (i_3, i_6, B, 2)
\]
 Additionally, the user-item interactions form edges:
\[
(u_1, i_1, t_1, 1, 1), \quad (u_1, i_2, t_2, 2, 1), \quad (u_1, i_3, t_3, 3, 1), \quad (u_1, i_4, t_4, 4, 1)
\]
The final graph integrates sequel-aware and user-item interactions, forming a heterogeneous graph:
\begin{equation}
\begin{split}
G &= \{(i_a, i_b, s, p_b) \mid i_a, i_b \in I_{SQ} \} \cup \{(u, i_{sa}, t, p_{iu}, 
p_{ui}) \\
&\mid i_{sa} \in I_{SA} \} \cup \{(u, i_{sq}, t, p_{iu}, p_{ui}) \mid i_{sq} \in I_{SQ} \}.
\end{split}
\end{equation}

\noindent where \( I_{SQ} \) represents sequel items and \( I_{SA} \) represents standalone items. For a given user sequence \( S_u = (i_1, i_2, \dots, i_k) \) with timestamps \( T_u = (t_1, t_2, \dots, t_k) \), predicting the next item involves identifying the most likely item connected to \( u \) in \( G_t^k \), leveraging both sequel-aware and temporal information.

\begin{figure}[h]
    \centering
    \begin{subfigure}[b]{0.25\textwidth}
         \centering
         \includegraphics[width=\textwidth]{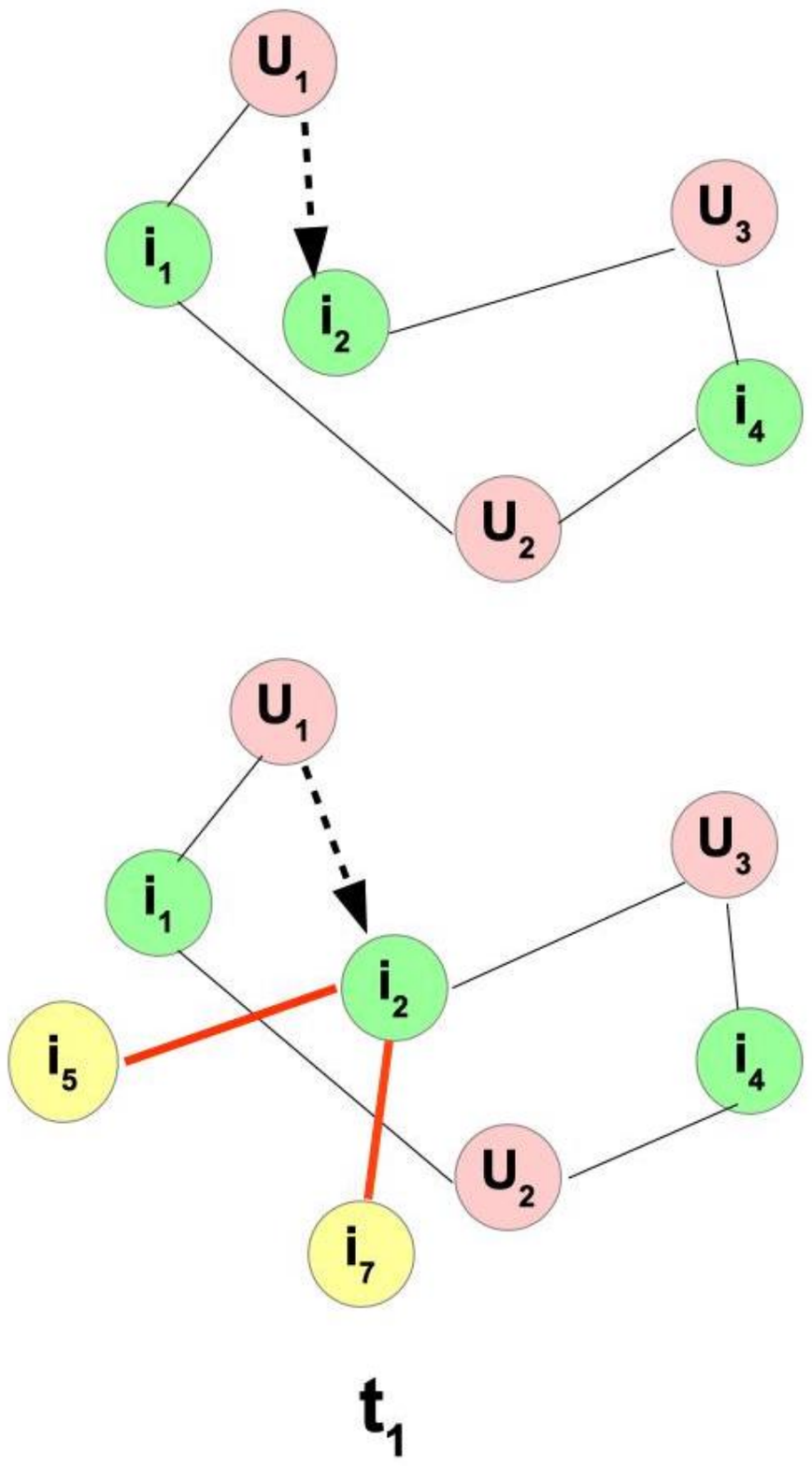}
     \end{subfigure}
     \hfill
     \begin{subfigure}[b]{0.25\textwidth}
         \centering
         \includegraphics[width=\textwidth]{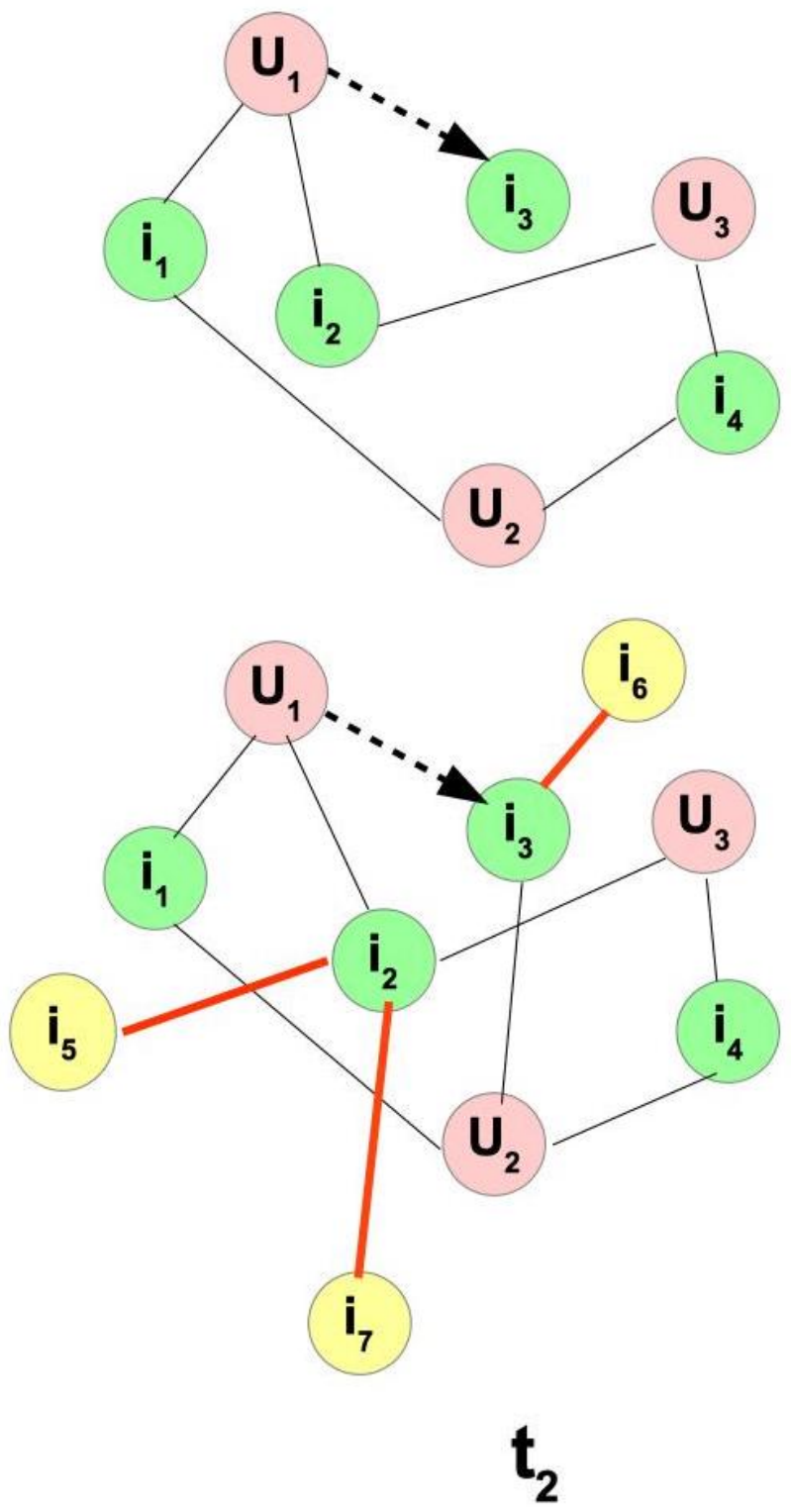}
     \end{subfigure}
     \hfill
     \begin{subfigure}[b]{0.25\textwidth}
         \centering
         \includegraphics[width=1\textwidth]{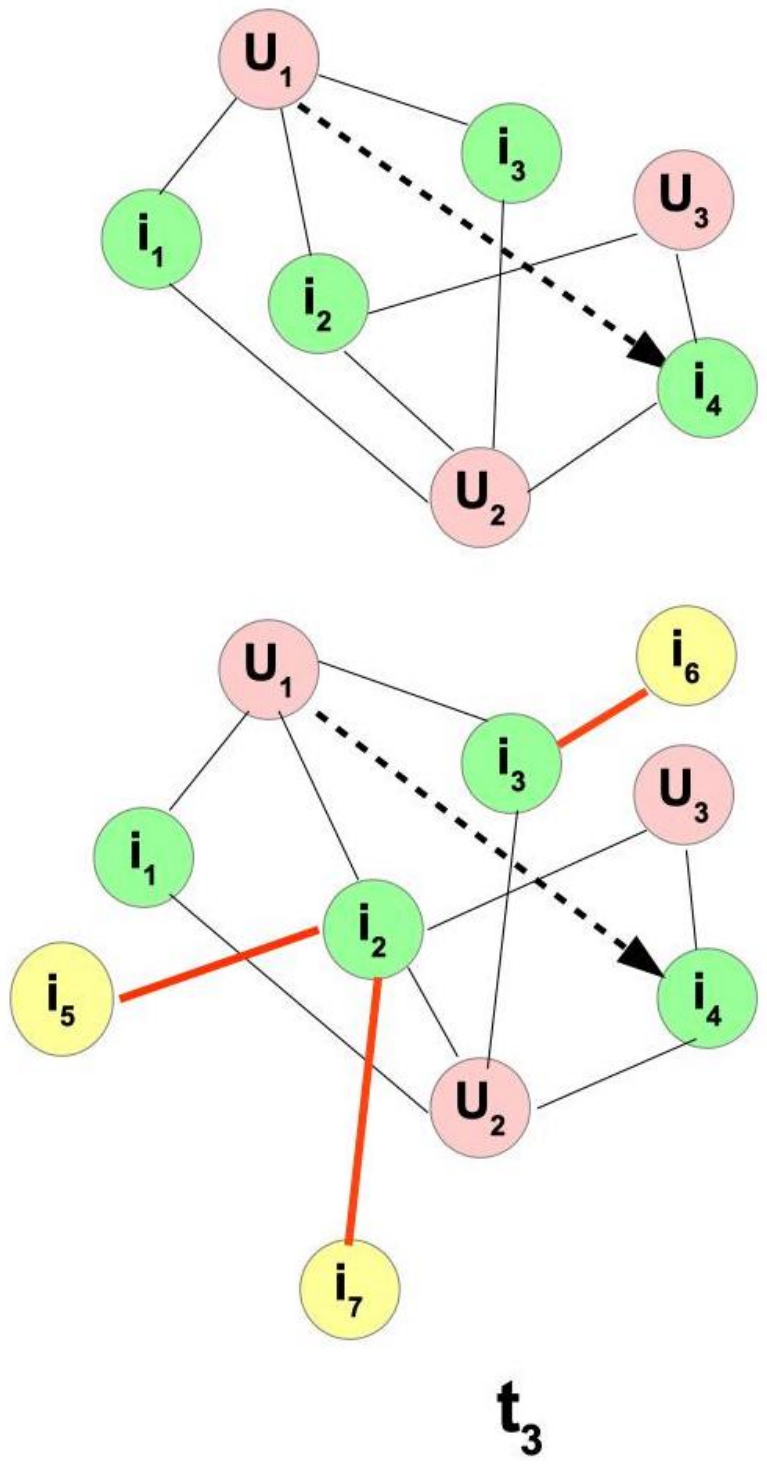}
     \end{subfigure}  
    \caption{Interaction of user-item graph at times $t_1, t_2, t_3$ for the illustrative example discussed in the text. Upper graphs illustrate traditional, dynamic sequential recommendation; Lower graphs illustrate sequel-aware, dynamic sequential recommendation. The graphs are heterogeneous at each time -- nodes belong to any of the following sets: $\mathcal{U}$, $i_{SA}$, or $i_{SQ}$; edges can be user-$i_{SA}$, user-${i_{SQ}}$, $i_{SA}$-$i_{SA}$, $i_{SQ}$-$i_{SQ}$, $i_{SA}$-$i_{SQ}$. }
    \label{illex2}
\end{figure}

\subsection{Sub-Graph Sampling}
\label{subsampling}
To improve efficiency and reduce noise, we employ a structured sub-graph sampling strategy that integrates both user-item interactions and sequel-based dependencies. Given a user node \( u \), we first select its most recent \( n \) first-order neighbors (historical interactions) from the graph \( G_t^k \). For each interacted item \( i \), we then sample the set of users who have interacted with \( i \), forming the second-order neighborhood. 

Our method further incorporates sequel-aware item-item connections. If an interacted item \( i \) belongs to a sequel series and is connected to subsequent items \( i_a \) and \( i_b \), these sequel items are also included in the sampled sub-graph. 
The final sub-graph is recursively expanded up to \( m \)-order neighborhoods, forming \( G_u^m(t_k) \), where \( m \) controls the depth of the sampled sub-graph.

\subsection{Heterogeneous Sequel-Aware Recommendation Network (HSAL-GNN)}

In this section, we introduce HSAL-GNN, a recommendation model that integrates user-item interactions with sequel-aware item dependencies to improve sequential learning. To achieve this, HSAL-GNN leverages a two-level information propagation mechanism:   
(1) sequel-aware item-item interactions, where items aggregate information from their sequels, allowing structured dependencies to be propagated throughout the network. 
(2) user-item interactions, where user and item representations are updated based on their historical interactions, capturing both long-term preferences and recent trends. 

\subsubsection{Graph Convolutional Neural Networks (GCN)}
Graph Convolutional Networks (GCNs) are widely used for learning node representations by aggregating information from neighboring nodes. The fundamental idea behind GCNs is to iteratively refine node embeddings through a message-passing mechanism. The user and item embeddings in a GCN-based model are updated as follows: $h_u^l = \frac{1}{|\mathcal{N}_u|} \sum_{i \in \mathcal{N}_u} \mathbf{W}_1^{(l-1)} h_i^{(l-1)}$
and $h_i^l = \frac{1}{|\mathcal{N}_i|} \sum_{u \in \mathcal{N}_i} \mathbf{W}_2^{(l-1)} h_u^{(l-1)},$ where $\mathbf{W}_1^{(l-1)}$ and $\mathbf{W}_2^{(l-1)} \in \mathbf{R}^{d \times d}$ are trainable transformation matrices for item and user embeddings at layer $l-1$, respectively. Here, $|\mathcal{N}_u|$ and $|\mathcal{N}_i|$ denote the number of neighboring nodes connected to user $u$ and item $i$.

\subsubsection{Sequel-Aware Message Propagation}

Sequel relationships capture structured dependencies between items within a series, enabling more effective item representation learning beyond user-item interactions. 
To incorporate structured sequel relationships, we introduce sequel-aware message propagation, where items aggregate information from their sequences.
For each item \( i \), we define its sequel-aware representation \( h_{i}^{\text{seq}} \), which is enriched by aggregating information from its subsequent sequel items \( \mathcal{N}_s(i) \), ensuring that sequential dependencies are explicitly modeled. The sequel-aware representation is computed as: $h_i^{\text{seq}}
 = \frac{1}{|\mathcal{N}_s(i)|} \sum_{j \in \mathcal{N}_s(i)} \left( h_j^{(l-1)} \cdot P(j) \right),$ where \( \mathcal{N}_s(i) \) represents the set of subsequent items in the sequel series. 

\subsubsection{User-Item Message Propagation}
To effectively model both long-term and short-term user preferences, we adopt a message propagation mechanism that integrates graph-based aggregation with sequence-based learning.

\noindent{\textbf{Long-Term Information}} 
Long-term user and item representations are learned by aggregating historical interactions using graph-based aggregation and sequential modeling. The user and item embeddings are updated as follows: $h_u^L = \sum_{i \in \mathcal{N}_u} \alpha_{ui} \left( \mathbf{W}_1 h_i + p_{r_{iu}}^V \right)$ and $h_i^L = \sum_{u \in \mathcal{N}_i} \beta_{iu} \left( \mathbf{W}_2 h_u + p_{r_{ui}}^V \right)$.
Here, \( p_{r_{iu}}^V \) and \( p_{r_{ui}}^V \) represent relative-order embeddings, and \( \alpha_{ui}, \beta_{iu} \) are attention weights computed via self-attention.

\noindent{\textbf{Short-Term Information}}
To model short-term user preferences and item trends, an attention mechanism is applied to dynamically weight the most recent interaction: $h_u^S = \sum_{i \in \mathcal{N}_u} \hat{\alpha}_{ui} h_i, \quad  
h_i^S = \sum_{u \in \mathcal{N}_i} \hat{\beta}_{iu} h_u.$

The attention scores \( \hat{\alpha}_{ui} \) and \( \hat{\beta}_{iu} \) assign higher importance to the latest interactions. By leveraging both \textbf{user-item interactions and sequel-aware dependencies}, our framework ensures that representations capture \textbf{temporal, structural, and sequel-based information} for improved sequential recommendation.

\subsubsection{Integrating Sequel-Aware Item Features with Long-Term Representations}
Once the item representations \( h_i^L \), \( h_i^S \), and the sequel-aware item representations \( h_i^{seq} \) are obtained, we integrate them to enhance the recommendation process. The fusion of these three representations is critical for capturing both individual item characteristics, and structured sequel relationships, ensuring that recommendations reflect both content-based dependencies and sequential progression across items. To achieve this, we explore four distinct fusion strategies, each offering a unique way to blend user-driven and sequel-aware information:

\noindent{\textbf{Summation:}} A straightforward yet effective method that assumes equal importance for user-item interactions and sequel-based item information. Before summation, we first employ an MLP-based transformation to aggregate long-term and short-term item representation into a unified representation: $ h_i^{\text{agg}} = \text{MLP}([h_i^L || h_i^{S}])$
Then, sequel-aware features are added: $h_i^{\text{sum}} = h_i^{\text{agg}} + h_i^{seq}$

\noindent{\textbf{Mean Aggregation:}} Averaging the two embeddings ensures a balanced contribution from both sources, preventing any dominant influence: $h_i^{\text{mean}} = \frac{1}{2} (h_i^{\text{agg}} + h_i^{seq})$

\noindent{\textbf{Concatenation:}} By concatenating long-term and short-term item representation with sequel-aware item representations, and applying a nonlinear transformation, this technique allows the model to learn \textit{higher-order interactions}: $h_i^{\text{concat}} = \text{ReLU} \left( \mathbf{W}_4 [h_i^L || h_i^S || h_i^{seq}] \right)$.

\noindent{\textbf{SeHGNN \cite{Yang_23a}-based Transformation:}} 
We employ SeHGNN, a heterogeneous graph-based transformation network, to adaptively fuse long-term, short-term, and sequel-aware item representations by capturing structural dependencies in a learnable manner, ensuring a context-aware representation: $ h_i^{SeHGNN} = SeHGNN([h_i^L, h_i^S, h_i^{seq}])$.

By leveraging multiple fusion techniques, our model effectively learns the optimal combination of item dependencies, ensuring robust and adaptive representations for sequential recommendations.

\subsubsection{Node Updating}
At this stage, we update the node representations by aggregating the long-term embedding, short-term embedding, sequel-aware representations, and the previous layer embedding to ensure both user interaction patterns and structured item dependencies are effectively captured.

\noindent\textbf{{Item node updates}}
For item nodes, the representation update from layer \( l-1 \) to layer \( l \) is computed as: $h_i^{(l)} = \tanh \left( \mathbf{W}_3^{(l)} \left[ h_i^{sum} \parallel h_i^{(l-1)} \right] \right), $
where $h_i^{sum}$ is the item representation obtained after combining long-term, short-term, and sequel-aware features earlier in the model.

\noindent\textbf{{User node updates}} User representations are updated as:\\
$h_u^{(l)} = \tanh \left( \mathbf{W}_4^{(l)} \left[ h_u^L \parallel h_u^S \parallel h_u^{(l-1)} \right] \right)$.

\subsubsection{Recommendation and Optimization}
To predict the next interaction for user \( u \), we formulate it as a link prediction task in the subgraph \( G_u^m(t_k) \). After applying \( L \) layers of our model, user embeddings from each layer are concatenated to form the final representation:

\begin{equation}
    \mathbf{h}_u = \mathbf{h}_u^{(0)} \parallel \mathbf{h}_u^{(1)} \parallel \dots \parallel \mathbf{h}_u^{(L)}.
\end{equation}

For a candidate item \( i \), the interaction score is computed as:

\begin{equation}
    s_{ui} = \mathbf{h}_u^\top \mathbf{W}_P \mathbf{e}_i,
\end{equation}
where \( \mathbf{W}_P \in \mathbb{R}^{(L+1)d \times d} \) is a trainable transformation matrix. The final recommendation scores are obtained using softmax normalization: $\hat{\mathbf{y}}_u = \text{softmax}(\mathbf{s}_u).$
Model parameters are optimized using binary cross-entropy loss with \( L_2 \) regularization: $Loss = - \sum_{S} \sum_{i=1}^{|I|} y_{ui} \log(\hat{y}_{ui}) + (1 - y_{ui}) \log(1 - \hat{y}_{ui}) + \lambda \| \Theta \|_2,$ where \( y_{ui} \) is the one-hot encoding vector of the ground truth items for the next interaction in sequence \( S^u \), \( \hat{y}_{ui} \) is the predicted probability of interaction, \( \Theta \) denotes all model parameters, \( \| \cdot \|_2 \) represents the \( L_2 \)-norm, \( \lambda \) controls the regularization strength to prevent overfitting.

\section{Experiments}
\label{exp}

In this section, we provide empirical results on three synthetic and three real-world datasets, to evaluate the effectiveness of HSAL-GNN. We answer the following research questions: (a) \textbf{RQ1:} How does HSAL-GNN perform compared to state-of-the-art sequential recommendation models? (b) \textbf{RQ2:} How effective is the sequel-aware item aggregation in improving recommendation accuracy compared to traditional user-item interaction-based models? (c) \textbf{RQ3:} What is the impact of different hyperparameter settings (number of HSAL-GNN layers, number of sequences, and maximum sequence length) on the model’s performance?


\subsection{Datasets}


\noindent{ \textbf{Synthetic Datasets: }} Three synthetic datasets (sequential ($I_{SQ})$, non-sequential ($I_{SA}$), and mixed ($I_{SQ} \cup I_{SA}$) items in the user-item sequence) are generated to test the performance of HSAL-GNN in a controlled setting. The key parameters used for generating the datasets are: (a) Number of users = 10,000 (b) Number of items = 500 (c) Items per user = 10 to 15 (d) Sequential items = 250 (e) Number of series = 20 to 30 (f) Max interactions/user = 15. 


%

\noindent{\textbf{Simulation Process:}} The simulation process assigns items to users based on a power-law distribution, capturing real-world popularity dynamics where some items are frequently interacted with while others are less common. Each user’s interactions are timestamped. We generate three datasets: Mixed (50\% sequels, 50\% standalone), Sequential (only sequels), and Standalone (only independent items).



\noindent{ \textbf{Real-World Datasets}} We evaluate our model on three real-world datasets: ML-1m, ML-100K \cite{ref_ML} and Goodreads (\cite{Wan_18a}, \cite{Wan_19a}), containing timestamped user-item interactions suitable for sequential recommendation. We took a sample of Goodreads dataset, and all reported results are based on sampled subsets. Goodreads explicitly provides series information, allowing direct incorporation of sequel-aware dependencies, while for MovieLens sequences were inferred by title matching (e.g., Movie X (Part 1), Movie X (Part 2)). After pre-processing, dataset statistics are summarized in Table~\ref{tab:dataset_stats}.

\begin{table}[h]
    \centering
    \caption{Dataset Statistics. Density = (number of interactions)/(number of users) * (number of items).}
    \begin{tabular}{l c c c c c c}
        \toprule
        \textbf{Datasets} & \textbf{\#users} & \textbf{\#items} & \textbf{\#series items} & \textbf{\#actions} & \textbf{Avg. length} & \textbf{Density} \\
        \midrule
       ML-1m & 6,040 & 3,900 & 312 & 1.0m & 163.5 & 4.79\% \\
       ML-100K & 943 & 1,682 & 67 & 100k & 106.0 & 6.30\% \\
        Goodreads-sampled & 2000 & 6474 & 1165 & 200k & 260.9 & 0.01\% \\
        \bottomrule
    \end{tabular}

    \label{tab:dataset_stats}
\end{table}

\subsection{Parameter Setup}
We implement HSAL-GNN using DGL and PyTorch\footnote{Code is available from URL removed for blind-review}. 
The embedding size and maximum sequence length are set to 50. The model is trained using the Adam optimizer (learning rate = 0.01, batch size = 50, weight decay $\lambda$ = 1e-4). Sub-graph sampling order $m$ is set to 4, and the number of GNN layers ($L$) is tuned per dataset. Experiments are run five times with different random seeds, and mean performance is reported. Baseline methods use default hyperparameters with embeddings set to 50 for consistency. All experiments are conducted on an HPC cluster with an NVIDIA A100 (40GB) GPU and an Intel Xeon E5-2660 v4 CPU.






\subsection{Baselines}  
We evaluate HSAL-GNN against three categories of sequential recommendation models: (1) conventional sequence models (GRU4Rec~\cite{Hidasi_2015}, SASRec~\cite{Kang_18a}, and BERT4Rec~\cite{Sun_19a}) leverage recurrent or Transformer-based architectures (2) uncertainty-aware and generative models (SVAE~\cite{Sachdeva_19a}, ACVAE~\cite{Xie_21a}, DiffuRec~\cite{Li_23a}) and (3) item-item interaction models (DGSR~\cite{Zhang_23a}, STOSA~\cite{Fan_22a}).

\subsection{Evaluation Settings}
We adopt the leave-one-out evaluation strategy, where the most recent user interaction is used for testing, the second most recent for validation, and the rest for training. To assess ranking quality, we use:
Hit@K, which measures how often the ground-truth item appears in the top-$K$ recommendations. 
NDCG@K, which prioritizes correctly ranked relevant items.

\begin{figure}[h]
    \centering
    \includegraphics[width=1\textwidth]{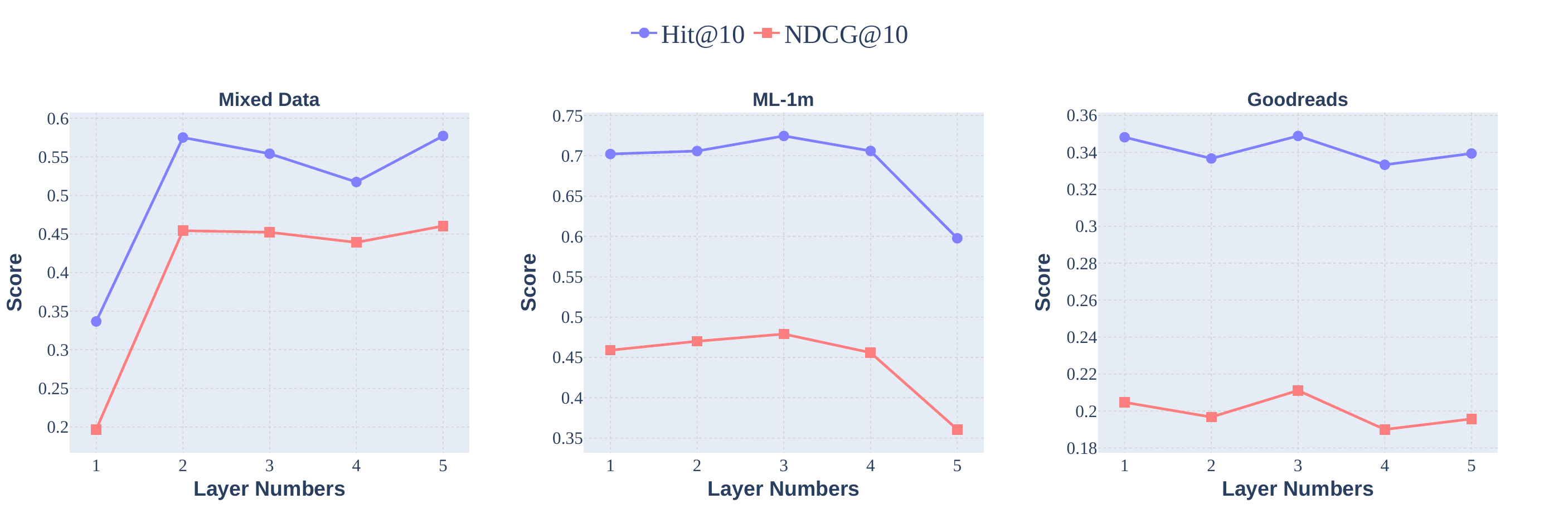}  
    \caption{Effect of propagation layer numbers on different datasets.}
    \label{fig:layer_analysis}
\end{figure}
\subsection{Performance Comparison (RQ1)} We evaluate HSAL-GNN against state-of-the-art baselines on synthetic and real-world datasets. The key observations from Table~\ref{tab:synthetic_results} and ~\ref{tab:realworld_results} are as follows:

\noindent \textbf{Synthetic Datasets: } In the Mixed dataset, where 50\% of the items belong to sequels, HSAL-GNN improves NDCG@10 by 5.80\% over the best baseline, showing its ability to effectively model both standalone and sequel items. In the Sequential dataset, where all items are part of structured sequels, HSAL-GNN is competitive to state-of-the-art models. In the Non-Sequential dataset, where items have no sequel relations, HSAL-GNN provides minimal improvements, confirming its primary advantage lies in sequel-aware learning.

\noindent \textbf{Real-World Datasets:} Goodreads (18\% sequel items) sees the highest gains, with HSAL-GNN improving NDCG@10 by 6.46\%, emphasizing the impact of sequel-aware aggregation. MovieLens 1m (ML-1m) (8\% sequel items) shows marginal improvements over DGSR, indicating that sequel modeling is beneficial but less pronounced in datasets with fewer sequels. MovieLens 100k (ML-100k) (4\% sequel items) sees little to no improvement, reinforcing that HSAL-GNN excels in datasets with substantial sequel content.
\textbf{Baseline Comparisons: }Traditional sequence models such as SASRec and BERT4Rec underperform as they ignore item-item relations. Graph-based DGSR is strong, but HSAL-GNN outperforms it in sequel-rich datasets by better leveraging structured dependencies.


\begin{figure}[h]
    \centering
    \includegraphics[width=1\textwidth]{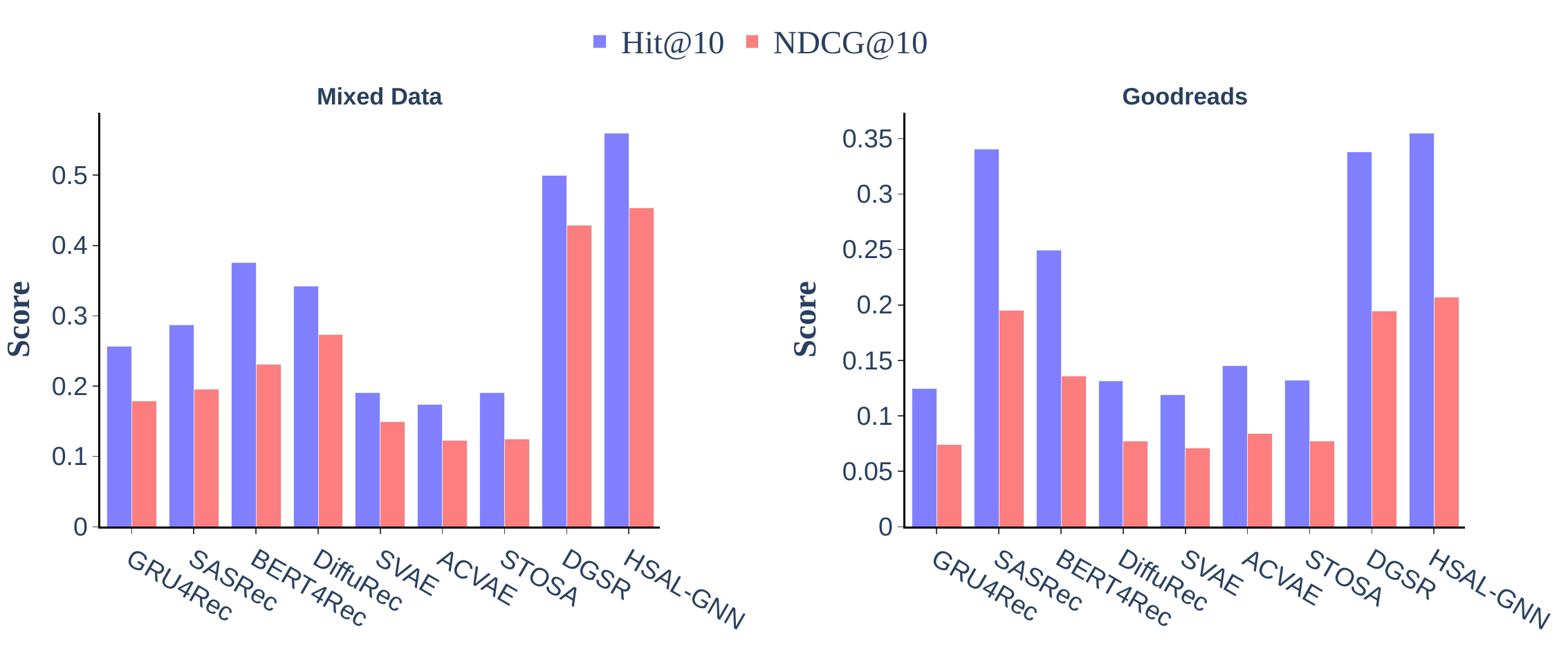}  
    \caption{Comparison of Hit@10 and NDCG@10 scores across different algorithms on the Mixed and Goodreads datasets}
    \label{fig:layer_analysis}
\end{figure}

\begin{figure}[h]
    \centering
    \begin{subfigure}{0.48\textwidth}
        \centering
        \includegraphics[width=\textwidth]{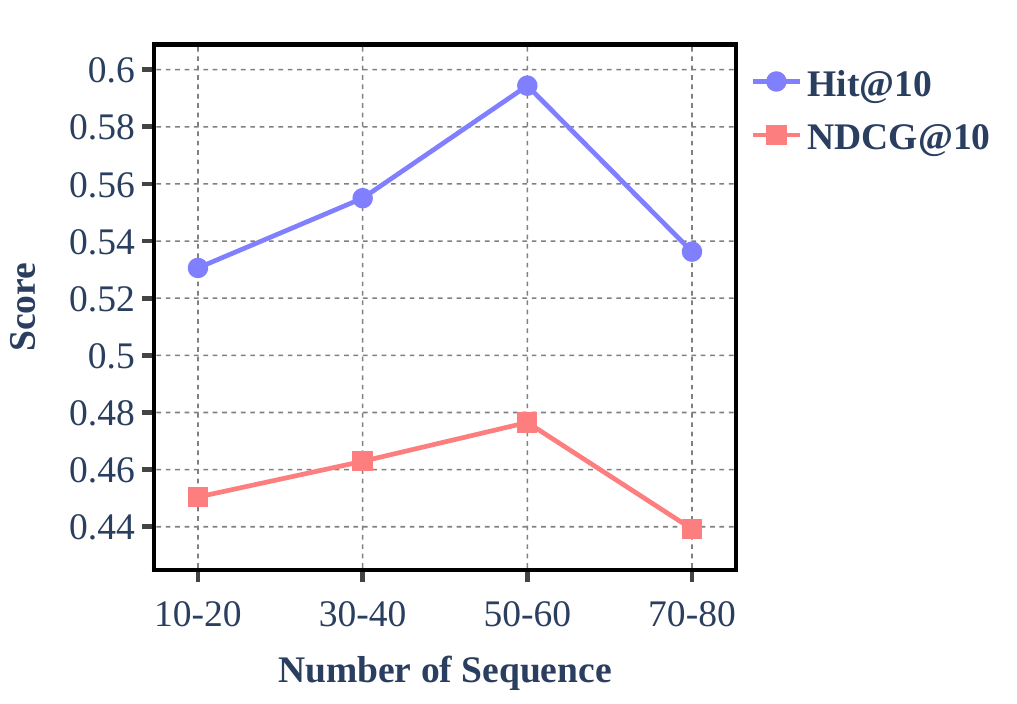}
        \label{fig:sequence_analysis}
    \end{subfigure}
    \hfill
    \begin{subfigure}{0.48\textwidth}
        \centering
        \includegraphics[width=\textwidth]{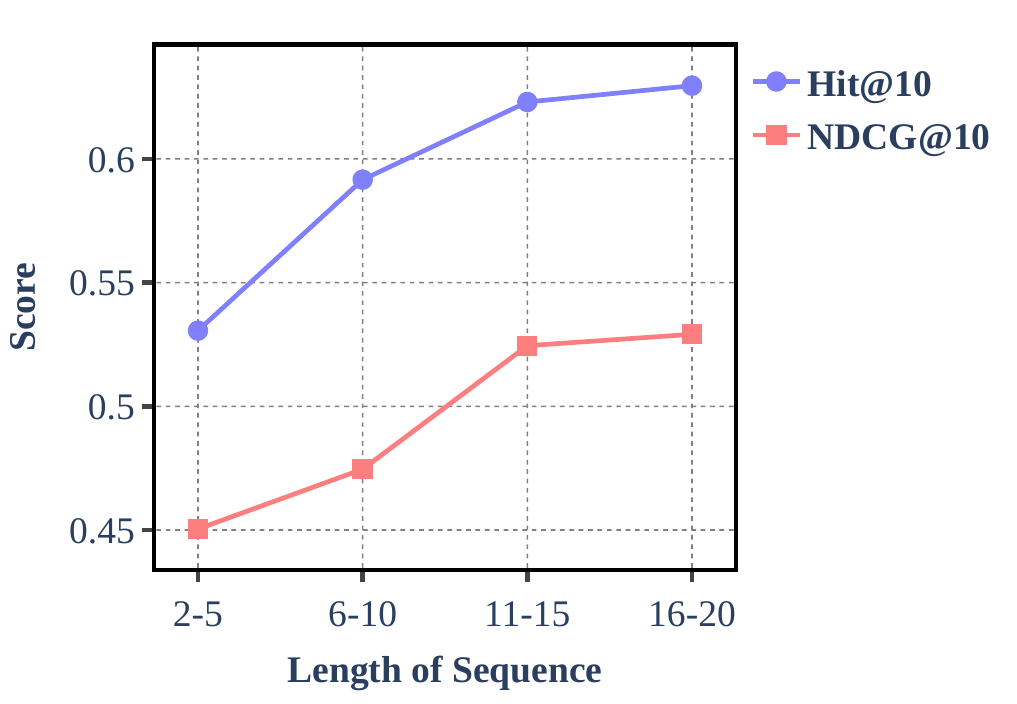}
        \label{fig:sequence_analysis}
        
    \end{subfigure}
    \caption{Effect of (a) number of sequence and (b) length of sequence}
    \label{fig:sequence_analysis}

\end{figure}

\begin{table}[h]
\centering
\renewcommand{\arraystretch}{1.2} 
\caption{\footnotesize \textbf{Performance Comparison on Synthetic Datasets}}
\label{tab:synthetic_results}
\resizebox{\textwidth}{!}{ 
\begin{tabular}{llccccccccc}
\toprule
\textbf{Dataset} & \textbf{Metric} & \textbf{GRU4Rec} & \textbf{SASRec} & \textbf{BERT4Rec} & \textbf{DiffuRec} & \textbf{SVAE} & \textbf{ACVAE} & \textbf{STOSA} & \textbf{DGSR} & \textbf{HSAL-GNN} \\ 
\midrule

\multirow{6}{*}{\textbf{Mixed}} 
& Hit@5   & 0.2321 & 0.2551 & 0.3267 & 0.3285 & 0.1816 & 0.1726 & 0.1819 & 0.4573 & \textbf{0.4891}  \\
& Hit@10  & 0.2563 & 0.2872 & 0.3758 & 0.3426 & 0.1911 & 0.1741 & 0.1908 & 0.4997 & \textbf{0.5596}  \\
& Hit@20  & 0.2954 & 0.3563 & 0.4366 & 0.3607 & 0.2097 & 0.1783 & 0.2096 & 0.5497 & \textbf{0.6157}  \\
& NDCG@5  & 0.1729 & 0.1857 & 0.2156 & 0.2696 & 0.1300 & 0.1191 & 0.1225 & 0.4153 & \textbf{0.4311}  \\
& NDCG@10 & 0.1794 & 0.1957 & 0.2313 & 0.2739 & 0.1497 & 0.1233 & 0.1252 & 0.4290 & \textbf{0.4539}  \\
& NDCG@20 & 0.1870 & 0.2115 & 0.2563 & 0.2882 & 0.1885 & 0.1281 & 0.1301 & 0.4410 & \textbf{0.4682}  \\ 
\midrule

\multirow{6}{*}{\textbf{Sequential}} 
& Hit@5   & 0.4176 & 0.7123 & 0.3966 & 0.7225 & 0.0222 & 0.0202 & 0.0382 & \textbf{0.9600}  & 0.9564  \\
& Hit@10  & 0.4355 & 0.7503 & 0.4561 & 0.7486 & 0.0344 & 0.0290 & 0.0586 & \textbf{0.9769}  & 0.9741  \\
& Hit@20  & 0.4732 & 0.7903 & 0.5581 & 0.7656 & 0.0612 & 0.0464 & 0.0781 & \textbf{0.9992}  & 0.9995  \\
& NDCG@5  & 0.3675 & 0.6708 & 0.3152 & 0.7303 & 0.0221 & 0.0305 & 0.0258 & \textbf{0.9498}  & 0.9476  \\
& NDCG@10 & 0.3744 & 0.6836 & 0.3567 & 0.7342 & 0.0381 & 0.0469 & 0.0311 & \textbf{0.9553}  & 0.9529  \\
& NDCG@20 & 0.3821 & 0.6938 & 0.3891 & 0.7385 & 0.0913 & 0.0748 & 0.0363 & \textbf{0.9607}  & 0.9592  \\ 
\midrule

\multirow{6}{*}{\textbf{Non-Sequential}} 
& Hit@5   & 0.0427 & 0.0836 & 0.0440 & 0.0433 & 0.0408 & 0.0453 & 0.0453 & \textbf{0.0529}  & 0.0519  \\
& Hit@10  & 0.0528 & 0.1337 & 0.0510 & 0.0508 & 0.0505 & 0.0567 & 0.0567 & \textbf{0.1043}  & 0.1001  \\
& Hit@20  & 0.0726 & 0.2420 & 0.0757 & 0.0707 & 0.0693 & 0.0785 & 0.0785 & \textbf{0.2056}  & 0.2009  \\
& NDCG@5  & 0.0301 & 0.0541 & 0.0307 & 0.0293 & 0.0322 & 0.0306 & 0.0306 & \textbf{0.0313}  & 0.0298  \\
& NDCG@10 & 0.0332 & 0.0692 & 0.0330 & 0.0316 & 0.0460 & 0.0342 & 0.0342 & \textbf{0.0476}  & 0.0432  \\
& NDCG@20 & 0.0381 & 0.0942 & 0.0386 & 0.0361 & 0.0808 & 0.0388 & 0.0388 & \textbf{0.0729}  & 0.0689  \\ 
\bottomrule
\end{tabular}
} 
\end{table}

\begin{table}[h]
\centering
\caption{\footnotesize \textbf{Performance Comparison on Real-World Datasets}}
\renewcommand{\arraystretch}{1.2} 

\label{tab:realworld_results}
\resizebox{\textwidth}{!}{ 
\begin{tabular}{llccccccccc} 
\toprule
\textbf{Dataset} & \textbf{Metric} & \textbf{GRU4Rec} & \textbf{SASRec} & \textbf{BERT4Rec} & \textbf{DiffuRec} & \textbf{SVAE} & \textbf{ACVAE} & \textbf{STOSA} & \textbf{DGSR} & \textbf{HSAL-GNN} \\
\midrule

\multirow{6}{*}{\textbf{ML-1m}} 
& Hit@5   & 0.2526 & 0.4408 & 0.2838 & 0.2441 & 0.2207 & 0.2331 & 0.2307 & 0.5905 & \textbf{0.5991} \\
& Hit@10  & 0.3145 & 0.6212 & 0.3713 & 0.3000 & 0.2595 & 0.2844 & 0.2691 & 0.7194 & \textbf{0.7260} \\
& Hit@20  & 0.3757 & 0.8310 & 0.4874 & 0.3453 & 0.2829 & 0.3258 & 0.2959 & 0.8269 & \textbf{0.8291} \\
& NDCG@5  & 0.1837 & 0.2990 & 0.2021 & 0.1781 & 0.1637 & 0.1715 & 0.1737 & 0.4348 & \textbf{0.4391} \\
& NDCG@10 & 0.2036 & 0.3569 & 0.2301 & 0.1960 & 0.1759 & 0.1879 & 0.1859 & 0.4764 & \textbf{0.4796} \\
& NDCG@20 & 0.2190 & 0.4098 & 0.2593 & 0.2074 & 0.1827 & 0.1983 & 0.1927 & 0.5034 & \textbf{0.5058} \\
\midrule

\multirow{6}{*}{\textbf{ML-100K}} 
& Hit@5   & 0.0465 & \textbf{0.6911} & 0.0923 & 0.01496 & 0.0037 & 0.1866 & 0.0174  & 0.5644  & 0.5493  \\
& Hit@10  & 0.0855 & \textbf{0.7919} & 0.1527 & 0.02662 & 0.0075 & 0.2715 & 0.0217  & 0.7288  & 0.7241  \\
& Hit@20  & 0.1471 & \textbf{0.9098} & 0.2577 & 0.05012 & 0.0119 & 0.3881 & 0.0323  & 0.8606  & 0.8483  \\
& NDCG@5  & 0.0290 & \textbf{0.5173} & 0.0604 & 0.00853 & 0.0041 & 0.1266 & 0.0153  & 0.3913  & 0.3819  \\
& NDCG@10 & 0.0415 & \textbf{0.5501} & 0.0799 & 0.01215 & 0.0093 & 0.1539 & 0.0166  & 0.4439  & 0.4347  \\
& NDCG@20 & 0.0569 & \textbf{0.5794} & 0.1062 & 0.01801 & 0.0158 & 0.1834 & 0.0193  & 0.4775  & 0.4679  \\
\midrule

\multirow{6}{*}{\textbf{Goodreads}} 
& Hit@5   & 0.0900 & 0.2389 & 0.1592 & 0.0940 & 0.0866 & 0.1038 & 0.0952 & 0.2412 & \textbf{0.2534} \\
& Hit@10  & 0.1250 & 0.3409 & 0.2497 & 0.1315 & 0.1190 & 0.1455 & 0.1322 & 0.3384 & \textbf{0.3549} \\
& Hit@20  & 0.1644 & 0.4699 & 0.3887 & 0.1704 & 0.1516 & 0.1894 & 0.1752 & 0.4632 & \textbf{0.4669} \\
& NDCG@5  & 0.0629 & 0.1628 & 0.1067 & 0.0652 & 0.0603 & 0.0707 & 0.0658 & 0.1638 & \textbf{0.1749} \\
& NDCG@10 & 0.0741 & 0.1954 & 0.1357 & 0.0773 & 0.0711 & 0.0842 & 0.0774 & 0.1948 & \textbf{0.2074} \\
& NDCG@20 & 0.0832 & 0.2268 & 0.1697 & 0.0864 & 0.0802 & 0.0945 & 0.0873 & 0.2262 & \textbf{0.2339} \\
\bottomrule
\end{tabular}
} 
\end{table}

\begin{table*}[h]
    \centering
    \caption{\footnotesize \textbf{Results of Ablation Experiments}}
    \renewcommand{\arraystretch}{1.2} 
    \label{tab:ablation_results}
    
    \resizebox{\textwidth}{!}{
    \begin{tabular}{llcccccc}
        \toprule
        \textbf{Ablation} & \textbf{Variants} & \multicolumn{2}{c}{\textbf{Synthetic Mixed Data}} & \multicolumn{2}{c}{\textbf{ML-1m}} & \multicolumn{2}{c}{\textbf{Goodreads}} \\
        \cmidrule(lr){3-4} \cmidrule(lr){5-6} \cmidrule(lr){7-8}
        & & \textbf{Hit@10} & \textbf{NDCG@10} & \textbf{Hit@10} & \textbf{NDCG@10} & \textbf{Hit@10} & \textbf{NDCG@10} \\
        \midrule
        \multirow{4}{*}{\textbf{Fusion Strategy}} 
        & Sum       & \textbf{0.5596} & \textbf{0.4540} & \textbf{0.7240} & \textbf{0.4795} & 0.3495 & 0.2042 \\
        & Mean      & 0.5276 & 0.4394 & 0.7189 & 0.4777 & 0.3439 & 0.2023 \\
        & Concat    & 0.5374 & 0.4408 & 0.7233 & 0.4809 & \textbf{0.3536} & \textbf{0.2089} \\
        & SeHGNN    & 0.5241 & 0.4324 & 0.7197 & 0.4755 & 0.3486 & 0.2044 \\
        \midrule
        \multirow{2}{*}{\textbf{Positional Embedding}} 
        & Sinusoidal & \textbf{0.5596} & \textbf{0.4539}  & \textbf{0.7241} & 0.4347 & \textbf{0.3549} & \textbf{0.2074} \\
        & Rotary     & 0.5086 & 0.4324  & 0.6864 & 0.4325 & 0.3455 & 0.2001 \\
        \bottomrule
    \end{tabular}
    }
\end{table*}

\subsection{Ablation Study (RQ2)}

To assess the contributions of different components in HSAL-GNN, we conduct an ablation study on the Synthetic Mixed dataset, ML-1m, and Goodreads. Table~\ref{tab:ablation_results} presents the results, revealing key insights:

\textbf{Fusion Strategy:} Among different aggregation methods, sum-based fusion achieves the highest performance across datasets, demonstrating its effectiveness in preserving structured sequel-aware information. 

\textbf{Positional Embedding:} Incorporating sinusoidal embeddings consistently outperforms the rotary embeddings, particularly in Goodreads and Synthetic Mixed datasets, reinforcing the importance of structured sequential encoding in sequel-aware recommendation.

\textbf{Overall Impact:} The ablation results highlight that sequel-aware modeling significantly contributes to recommendation quality, with structured fusion and effective positional encoding playing crucial roles. The superior performance of HSAL-GNN demonstrates its ability to capture both user-item interactions and sequel-based dependencies, making it particularly effective in sequel-rich datasets.
\subsection{Sensitivity Analysis of Hyperparameters (RQ3)} To investigate the impact of key hyperparameters on model performance, we examine the number of sequences, sequence length, and the number of HSAL-GNN layers.

\noindent \textbf{Effect of Number of Sequences and Sequence Length}
We analyze the impact of varying the number of sequences and sequence length on recommendation performance. As seen in Figure~\ref{fig:sequence_analysis} $(a)$, increasing the number of sequences initially improves Hit@10 and NDCG@10 scores, but performance declines beyond a certain point. A similar trend is observed with sequence length in Figure~\ref{fig:sequence_analysis} $(b)$, where longer sequences contribute to better performance until saturation is reached. This indicates that leveraging sequential dependencies enhances recommendations, but excessive sequence length may introduce noise.

\noindent \textbf{Effect of HSAL-GNN Layers}
We further study the role of HSAL-GNN layers across different datasets, including Mixed Data, ML-1m, and Goodreads, as illustrated in Figure~\ref{fig:layer_analysis}. The results show that increasing the number of layers improves Hit@10 and NDCG@10 scores up to a certain depth, beyond which performance starts to decline. This suggests that deeper layers facilitate better user-item interaction modeling, but excessive depth may lead to over-smoothing, reducing effectiveness.
These findings highlight the importance of balancing sequence length and network depth to achieve optimal recommendation performance.


\section{Conclusion}
\label{conc}
This paper presents a heterogeneous sequel-aware Graph Neural Network for sequential learning. While literature on sequential recommendation has focused extensively on user-item sequences, their representation and dynamic behavior, this work shows that if users interact with items that occur in a sequence and this information is incorporated in the design of algorithms, it improves recommendation substantially -- particularly if user-item sequences are a mix of standalone items and those with sequences. Extensive empirical results on three variants of synthetic datasets and three, large, real-world datasets show that the amount of sequel information already present in the data plays a crucial role in determining the performance of HSAL-GNN. Future work involves deployment in real-world settings and exploration of the performance of scalable variants of the algorithm designed.


%
%

\begin{thebibliography}{88}
\bibitem{Hidasi_2015}
Hidasi, B., Karatzoglou, A., Baltrunas, L. and Tikk, D., ``Session-based recommendations with recurrent neural networks". arXiv preprint arXiv:1511.06939, 2015.

\bibitem{Zhang_23a}
Mengqi Zhang, Shu Wu, Xueli Yu, Qiang Liu, and Liang Wang, ``Dynamic Graph Neural Networks for Sequential Recommendation", IEEE Trans. on Knowl. and Data Eng. 35 (5), 4741–4753,2023. 

\bibitem{Li_23a}
Zihao Li, Aixin Sun, and Chenliang Li, ``DiffuRec: A Diffusion Model for Sequential Recommendation", ACM Trans. Inf. Syst. 42 (3),1--28, 2023.

\bibitem{Fan_21a}
Ziwei Fan, Zhiwei Liu, Shen Wang, Lei Zheng, and Philip S. Yu, ``Modeling Sequences as Distributions with Uncertainty for Sequential Recommendation", In Proceedings of the  International Conference on Information and Knowledge Management (CIKM), pp 3019–3023, 2021. 

\bibitem{Fan_22}
Ziwei Fan, Zhiwei Liu, Yu Wang, Alice Wang, Zahra Nazari, Lei Zheng, Hao Peng, and Philip S. Yu, ``Sequential Recommendation via Stochastic Self-Attention", In Proceedings of the ACM Web Conference (WWW '22). Association for Computing Machinery, New York, NY, USA, 2036–2047, 2022. 

\bibitem{Rendle_10a}
S. Rendle, C. Freudenthaler, and L. Schmidt-Thieme, ``Factorizing personalized Markov Chains for next-basket recommendation", In Proceedings of the International conference on World Wide Web, pp. 811–820, 2010.

\bibitem{Quadrana_17}
M. Quadrana, A. Karatzoglou, B. Hidasi, and P. Cremonesi, ``Personalizing session-based recommendations with hierarchical recurrent neural networks", In Proceedings of the Eleventh ACM Conference on Recommender Systems, pp. 130–137, 2017.

\bibitem{Hidasi_18a}
B. Hidasi and A. Karatzoglou, ``Recurrent neural networks with top-k gains for session-based recommendations", In Proceedings of the 27th ACM International Conference on Information and Knowledge Management, 2018, pp. 843–852.

\bibitem{Avatar}
Avatar, https://en.wikipedia.org/wiki/Avatar

\bibitem{Sak_14a}
H. Sak, A. W. Senior, and F. Beaufays, ``Long short-term memory recurrent neural network architectures for large scale acoustic modeling", 2014.

\bibitem{Chung_14a}
J. Chung, C. Gulcehre, K. Cho, and Y. Bengio, ``Empirical evaluation of gated recurrent neural networks on sequence modeling", arXiv preprint arXiv:1412.3555, 2014.

\bibitem{Tang_18a}
J. Tang and K. Wang, ``Personalized top-n sequential recommendation via convolutional sequence embedding", In Proceedings of the Eleventh ACM International Conference on Web Search and Data Mining, 2018, pp. 565–573.

\bibitem{Kang_18a}
W.-C. Kang and J. McAuley, ``Self-attentive sequential recommendation", In IEEE International Conference on Data Mining(ICDM). IEEE, pp. 197–206, 2018.

\bibitem{Liu_18a}
Q. Liu, Y. Zeng, R. Mokhosi, and H. Zhang, ``Stamp: short-term attention/memory priority model for session-based recommendation", In Proceedings of the 24th ACM SIGKDD International Conference on Knowledge Discovery and Data Mining, 2018, pp. 1831–1839.

\bibitem{Liang_18a}
D. Liang, R. G. Krishnan, M. D. Hoffman, and T. Jebara, ``Variational autoencoders for collaborative filtering", In Proceedings of the World Wide Web Conference, pp 689–698, 2018.

\bibitem{Sachdeva_19a}
N. Sachdeva, G. Manco, E. Ritacco, and V. Pudi, ``Sequential variational autoencoders for
collaborative filtering", In Proceedings of the ACM International Conference on Web Search and Data Mining, pp 600–608, 2019.

\bibitem{Xie_21a}
Z. Xie, C. Liu, Y. Zhang, H. Lu, D. Wang, and Y. Ding, ``Adversarial and contrastive variational autoencoder for sequential recommendation", In Proceedings of the Web Conference, pp 449–459, 2021.

\bibitem{Shani_05a}
Guy Shani, David Heckerman, Ronen I. Brafman, and Craig Boutilier, ``An MDP-based recommender system", Journal of Machine Learning Research, 6, 9, 1265–1295, 2005.

\bibitem{Sun_19a}
F. Sun, J. Liu, J. Wu, C. Pei, X. Lin, W. Ou, and P. Jiang, ``BERT4Rec: Sequential recommendation with bidirectional encoder representations from Transformer", In Proceedings of the ACM International Conference on Information and Knowledge Management, pp 1441–1450, 2019.

\bibitem{Ding_21a}
Y. Ding, Y. Ma, W. K. Wong, and T.S. Chua, ``Leveraging two types of global graph for sequential fashion recommendation", In Proceedings of the International Conference on Multimedia Retrieval. pp 73–81, 2021.

\bibitem{Wu_19a}
S. Wu, Y. Tang, Y. Zhu, L. Wang, X. Xie, and T. Tan, ``Session-based recommendation with graph neural networks", In Proceedings of the AAAI Conference on Artificial Intelligence, Vol. 33, pp 346–353, 2019.

\bibitem{Scarselli_09a}
F. Scarselli, M. Gori, A. C. Tsoi, M. Hagenbuchner and G. Monfardini, ``The Graph Neural Network Model," in IEEE Transactions on Neural Networks, vol. 20, no. 1, pp. 61-80, Jan. 2009.

\bibitem{Yang_23a}
Yang, X., Yan, M., Pan, S., Ye, X. and Fan, D., 2023, June. Simple and efficient heterogeneous graph neural network. In Proceedings of the AAAI conference on artificial intelligence (Vol. 37, No. 9, pp. 10816-10824).

\bibitem{Kang_18a}
Kang, W.C. and McAuley, J., 2018, November. Self-attentive sequential recommendation. In 2018 IEEE international conference on data mining (ICDM) (pp. 197-206). IEEE.

\bibitem{Li_23a}
Li, Z., Sun, A. and Li, C., 2023. Diffurec: A diffusion model for sequential recommendation. ACM Transactions on Information Systems, 42(3), pp.1-28.

\bibitem{Fan_22a}
Fan, Z., Liu, Z., Wang, Y., Wang, A., Nazari, Z., Zheng, L., Peng, H. and Yu, P.S., 2022, April. Sequential recommendation via stochastic self-attention. In Proceedings of the ACM web conference 2022 (pp. 2036-2047).

\bibitem{Wan_18a}
Wan, M. and McAuley, J. ``Item Recommendation on Monotonic Behavior Chains", RecSys, 2018. 

\bibitem{Wan_19a}
Wan, M., Misra, R., Nakashole, N., McAuley, J., ``Fine-Grained Spoiler Detection from Large-Scale Review Corpora", ACL, 2019.

\bibitem{ref_ML}
MovieLens, \url{https://grouplens.org/datasets/movielens/}


\end{thebibliography}
%

\end{document}


\title{Heterogeneous Sequel-Aware Graph Neural Networks for Sequential Learning}

\titlerunning{HSAL-GNN}

\author{Author information scrubbed for double-blind reviewing}



\maketitle              
\appendix
\section{Appendix A}
\label{appendA}

\begin{figure}[!h]
    \centering
    \includegraphics[width=1\textwidth]{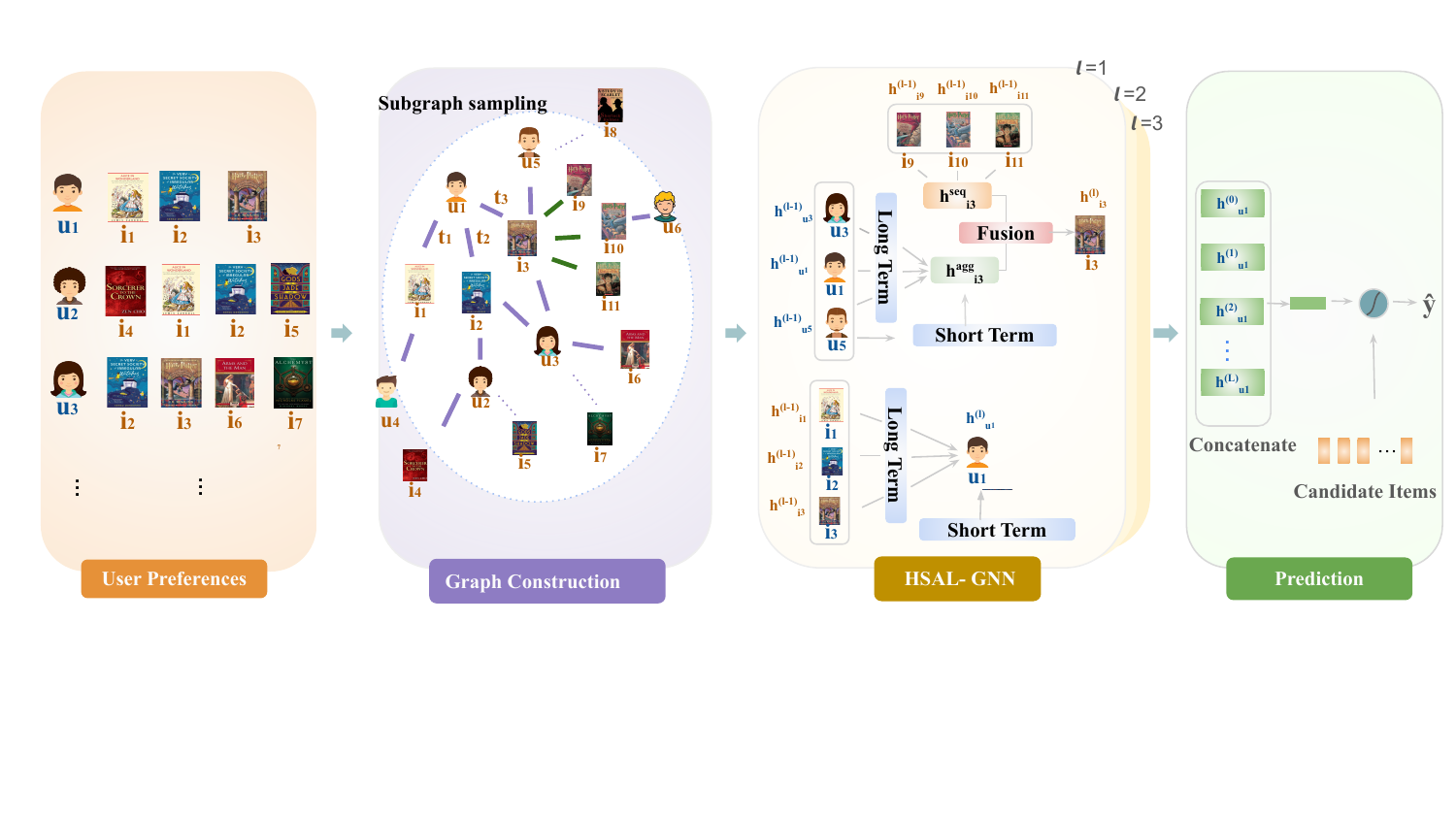}
    \caption{Overview of HSAL-GNN.}
    \label{fig:hsal_gnn_overview}
\end{figure}

Figure~\ref{fig:hsal_gnn_overview} presents the overall framework for HSAL-GNN. While all the details have been described in the main paper, the appendix presents a visual representation of the algorithm framework. Also, a schematic view of the sub-sampling algorithm and the forward propagation of HSAL-GNN (described in Section 4) are presented here.
\begin{table*}[t]

\centering
\begin{minipage}{0.49\textwidth}
    \begin{algorithm}[H]
    \RaggedRight
    \small

        \caption{Sequel-Aware Sub-graph Sampling Algorithm}
        \label{alg:sequel_subgraph_sampling}
        \textbf{Input}: Sequence $S^u = (i_1, i_2, \dots, i_k)$, time sequence $T^u = (t_1, t_2, \dots, t_k)$, dynamic graph $G^{t_k}$, and the order of sub-graph $m$. \\
        \textbf{Output}: The $m$-order sub-graph $G_u^m(t_k)$.
        \begin{algorithmic}[1]
            \STATE // Initialization
            \STATE $U_m, U_{\text{temp}} \gets \{u\}$, $I_m, I_{\text{temp}} \gets \{i_1, \dots, i_k\}$, $j \gets 0$
            \STATE Partition $I_{\text{temp}}$ into $I_{\text{SA}}$ and $I_{\text{SQ}}$ 
            \FOR{$i \in I_{\text{SQ}}$}
                \STATE $I_{\text{temp}} \gets I_{\text{temp}} \cup \text{Sequel}(i) \setminus I_{\text{temp}}$ 
            \ENDFOR
            \STATE // Node sampling
            \WHILE{$j \leq m$}
                \FOR{$i \in I_{\text{temp}}$}
                    \STATE $U_{\text{temp}} \gets U_{\text{temp}} \cup \mathcal{N}_i$
                \ENDFOR
                \STATE $U_{\text{temp}} \gets U_{\text{temp}} \setminus U_m$
                \STATE $U_m \gets U_m \cup U_{\text{temp}}$
                \IF{$U_{\text{temp}} = \emptyset$}
                    \STATE \textbf{break}
                \ENDIF
                \FOR{$u \in U_{\text{temp}}$}
                    \STATE $I_{\text{temp}} \gets I_{\text{temp}} \cup \mathcal{N}_u$            
                    \STATE Partition $\mathcal{N}_u$ into $I_{\text{SA}}$ and $I_{\text{SQ}}$
                    \FOR{$i \in I_{\text{SQ}}$}
                        \STATE $I_{\text{temp}} \gets I_{\text{temp}} \cup \text{Sequel}(i) \setminus I_{\text{temp}}$
                    \ENDFOR
                \ENDFOR
                \STATE $I_{\text{temp}} \gets I_{\text{temp}} \setminus I_m$
                \STATE $I_m \gets I_m \cup I_{\text{temp}}$
                \IF{$I_{\text{temp}} = \emptyset$}
                    \STATE \textbf{break}
                \ENDIF
                \STATE $j \gets j + 1$
            \ENDWHILE
            \STATE // Sub-graph generation
            \STATE $G_u^m(t_k) = (U_m, I_m), U_m, I_m \in G^{t_k}$
        \end{algorithmic}
    \end{algorithm}
\end{minipage}
\hfill
\begin{minipage}{0.49\textwidth}
    \begin{algorithm}[H]
    \RaggedRight
    \small

        \caption{The HSAL-GNN framework (forward propagation)}
        \label{alg:HSAL-GNN}
        Input: Sequence $S^u = (i_1, i_2, \cdots , i_k)$, timestamp sequence $T^u = (t_1, t_2, \cdots , t_k)$, all sequences of users, and GNN layer number $L$.\\
        Output: The next item $i_{k+1}$ of $S^u$.
        \begin{algorithmic}[1] 
            \STATE // Sequel-Aware Graph Construction
            \STATE Convert all user sequences into a sequel-aware graph $G$
            \STATE // Sub-graph generation for $S^u$
            \STATE Run Algorithm 1 to generate $G_u^m (t_k)$ from $G^{t_k}$
            \STATE // Initialization of node representations
            \STATE $h_u^{(0)} \leftarrow e_u, \quad h_i^{(0)} \leftarrow e_i, \quad \forall u, i \in G_u^m(t_k)$
            \STATE // Representation update using HSAL-GNN
            \FOR{$l \in [1 : L]$}
                \STATE $h_u^{(l)}, h_i^{(l)} \leftarrow \text{HSAL}(h_u^{(l-1)}, h_i^{(l-1)}, G_u^m(t_k))$
                \STATE $h_u^{(L)}, h_i^{(L)} \leftarrow$ Long-term Information 
                \STATE  $h_u^{(S)}, h_i^{(S)} \leftarrow$ Short-term Information 

                \STATE $h_i^{\text{seq},(l-1)} = \frac{1}{|\mathcal{N}_s(i)|} \sum\limits_{j \in \mathcal{N}_s(i)} \left( h_j^{(l-1)} \cdot P(j) \right)$

                \STATE $h_i^{\text{agg}} = \text{MLP} \left(  \left[ h_i^{(L)} \Vert h_i^{(S)} \right] \right)$
                
                \STATE $h_i^{\text{fused}} = \text{SUM} \left(h_i^{\text{agg}}, h_i^{\text{seq},(l-1)} \right)$
                \STATE Compute final item embedding:
                \STATE $h_i^{(l)} \leftarrow \tanh \left( \mathbf{W}_3^{(l)} \left[ h_i^{\text{fused}} \Vert h_i^{(l-1)} \right] \right)$
                \STATE Compute final user embedding:
                \STATE $h_u^{(l)} \leftarrow \tanh \left( \mathbf{W}_4^{(l)} \left[ h_u^{L} \Vert h_u^{S} \Vert h_u^{(l-1)} \right] \right)$
            \ENDFOR
            \STATE // Prediction of next item
            \STATE $h_u = h_u^{(0)} \, \Vert \, h_u^{(1)} \, , \cdots , \Vert \, h_u^{(L)}$
            \STATE Next item $\leftarrow \arg\max_{i \in V}(h_u^\top W_P e_i)$
        \end{algorithmic}
    \end{algorithm}
\end{minipage}
\end{table*}


